\documentclass[prb,aps,twocolumn,superscriptaddress,10pt]{revtex4-2}

\usepackage{graphicx}

\usepackage{mathtools}

\usepackage{amsmath}

\usepackage{mathrsfs}

\usepackage{amssymb}

\usepackage{bbold}

\usepackage{bm}

\usepackage[dvipsnames]{xcolor}

\usepackage[
	colorlinks=true,
	linkcolor=blue,
	anchorcolor=blue,
	citecolor=blue,
	urlcolor=blue
]{hyperref}
\usepackage{wasysym}
\DeclareSymbolFont{usualmathcal}{OMS}{cmsy}{m}{n}
\DeclareSymbolFontAlphabet{\mathcal}{usualmathcal}
\usepackage{braket}

\makeatletter

\newcommand{\be}{\begin{equation}}
\newcommand{\ee}{\end{equation}}
\newcommand{\bea}{\begin{eqnarray}}
\newcommand{\eea}{\end{eqnarray}}

\newcommand{\mc}{\mathcal}
\newcommand{\mb}{\mathbf}

\begin{document}
\title{Majorana Fermi surface state in a network of quantum spin chains}

\author{Fabrizio G. Oliviero}
\affiliation{Departamento de F\'isica Te\'orica e Experimental, Universidade
Federal do Rio Grande do Norte, 59072-970 Natal-RN, Brazil}
\author{Weslei B. Fontana}
\affiliation{International Institute of Physics, Universidade Federal do Rio Grande
do Norte, 59078-970 Natal-RN, Brazil}
\author{Rodrigo G. Pereira}
\affiliation{Departamento de F\'isica Te\'orica e Experimental, Universidade
Federal do Rio Grande do Norte, 59072-970 Natal-RN, Brazil}
\affiliation{International Institute of Physics, Universidade Federal do Rio Grande
do Norte, 59078-970 Natal-RN, Brazil}

\begin{abstract}
We use junctions of critical spin-1 chains as the basic elements to construct a honeycomb network that harbors a gapless chiral spin liquid phase. The low-energy modes   are described by  spin-1 Majorana fermions  that  form a two-dimensional Fermi surface when the interactions at the junctions are tuned to the vicinity of chiral fixed points with staggered chirality. We discuss the physical properties and the stability of this chiral spin liquid phase against perturbations from the point of view of the effective field theory for the network. We find clear connections with the excitation spectrum obtained in parton constructions  on the kagome lattice. 

\end{abstract}
\maketitle

\section{Introduction}

Mott insulators can display exotic quantum phases in which  spin  fractionalization gives rise to low-energy  fermionic excitations  \cite{sachdev2023quantum}. In materials regarded as quantum spin liquid  candidates \cite{SavaryIOP2017,Knolle2019,Broholm2020}, the observation of a constant magnetic susceptibility, linear specific heat and linear thermal conductivity at low temperatures is often interpreted as evidence for a Fermi surface of fractionalized excitations. Within  effective  theory descriptions \cite{Lee2006}, these gapless  fermionic modes are strongly coupled to emergent gauge fields, and one may question whether these   phases remain stable when interactions are treated beyond the mean-field level  \cite{WenPRB2002,WenPRB2004,Lee2008}. 


 Quantum spin liquids with Fermi surfaces  become more stable when   the gauge structure is discrete and  time reversal  and inversion symmetries are broken \cite{Barkeshli2013}. As opposed to a U(1) spin liquid, whose gapless photon-like modes   mediate long-range interactions between fermions, a $\mathbb Z_2$ spin liquid has gapped vortex-like excitations known as visons \cite{Senthil2000}. Gauge-field fluctuations can be safely neglected  in the limit where visons have a large gap and a small effective bandwidth. In addition, breaking  time reversal and inversion symmetries, as in chiral spin liquids (CSLs) \cite{BieriPRB2016},   protects the Fermi surface against   pairing instabilities    \cite{Barkeshli2013,Metlitski2015}. In fact, gapless CSL phases have been found in numerical studies of  lattice models where time reversal symmetry is broken either spontaneously  or by three-spin interactions \cite{Gong2011,BauerPRB2019,OlivieroSciPost2022}. In these cases, the formation of the Fermi surface is associated with a staggered scalar spin chirality on frustrated lattices. Moreover, there are examples of exactly solvable models  where   spins fractionalize into Majorana fermions and static $\mathbb Z_2$ gauge fields, and the Majorana fermions form a stable Fermi surface  \cite{ChuaPRB2011,MotrunichPRB2011a,MotrunichPRB2011b,Hermanns2014,Chari2021}.

In this work we present an analytical approach that employs  quantum  spin chains coupled   by time-reversal-symmetry-breaking interactions as  building blocks of a Majorana Fermi surface state. Arrays of one-dimensional (1D) systems have been shown to realize both gapped \cite{SelaPRB2015,ThomalePRB2015,Huang2016,Patel2016,Lecheminant2017,Leviatan2020} and gapless \cite{PereiraScipost2018,Slagle2022} spin liquids. Such coupled-wire constructions usually hinge on the  assumption of a renormalization group (RG) flow of judiciously selected  interchain interactions to strong coupling. By contrast, here we start from junctions of spin chains with boundary interactions tuned to a chiral fixed point \cite{OshikawaIOP2006,BuccheriPRB2018,BuccheriNPB2019,XavierPRB2022}. When the spin chains are coupled to form a network with uniform spin chirality at the junctions, this approach  leads to gapped CSLs with Abelian \cite{FerrazPRL2019} or non-Abelian \cite{XavierScipost2023}  topological order. Our goal here is to show that the same approach applied to a network with staggered spin chirality describes  a gapless CSL with a Fermi surface   descended from the chiral 1D modes.

To construct a $\mathbb Z_2$ CSL with a Majorana Fermi surface, we consider a network of critical spin-1 chains described by the SU(2)$_2$ Wess-Zumino-Novikov-Witten (WZNW)  model \cite{Gogolin2004}. The latter is a conformal field theory (CFT) with central charge $c=3/2$ and  admits a representation in terms of three  Majorana fermions for each chain \cite{Tsvelik1990,Allen2000}. The conditions for reaching the chiral fixed point of a junction of three spin-1 chains were discussed in Ref. \cite{XavierPRB2022}. Imposing   a   staggered chirality pattern on the junctions forming a honeycomb network, we show that the low-energy excitations of the system are chiral Majorana modes that run along three  zigzag directions in the network. We then consider the leading perturbations allowed by symmetry when the model parameters  deviate from the chiral fixed point. The theory contains a marginal   operator that introduces backscattering of Majorana fermions at the junctions and can be treated exactly. This operator turns the fermionic spectrum into an authentic 2D dispersion with a line Fermi surface, closely related  to that  obtained in parton mean-field theories with Majorana fermions on the kagome and triangular lattices \cite{BauerPRB2019,SachdevPRB2011}. We then analyze the effects of the operator associated with  the spin-1/2 primary field of the SU(2)$_2$ WZNW model. While this perturbation is highly relevant at the 1D fixed point, we show that  deep in the 2D regime  this operator governs the dynamics of gapped vison excitations, thus becoming  irrelevant at low energies. Therefore,  this network approach provides a path to tame the gauge-field fluctuations and stabilize an SU(2)-invariant  gapless CSL without resorting to    mean-field approximations.




This paper is organized as follows. In Sec. \ref{Yjunctionsec}, we review the basic aspects of the critical spin chains that constitute the Y junction. In Sec. \ref{section-3}, we show how to obtain the network model by suitably coupling the junctions tuned to chiral fixed points. In Sec. \ref{section 4}, we discuss the stability of the CSL phase against  perturbations that modify the  Majorana fermion spectrum and create visons.  In Sec. \ref{section 5}, we examine the effects of turning on a magnetic field in the gapless CSL phase. Final remarks and possible directions for future work are presented in Sec. \ref{conclusions}.

\section{Junction of   spin-1 chains \label{Yjunctionsec}}

Let us briefly review the   theory for  a single Y junction of three critical spin-1 chains \cite{XavierPRB2022}. The lattice model is given by  $H_{Y}=H_c+H_{\mc B}$, where $H_c$  contains the intrachain  interactions: 
\begin{equation}
H_{c}=J\sum_{\alpha=1}^{3}\sum_{j=1}^L\left[\mathbf{S}_{j,\alpha}\cdot\mathbf{S}_{j+1,\alpha}-\left(\mathbf{S}_{j,\alpha}\cdot\mathbf{S}_{j+1,\alpha}\right)^{2}\right]\,,\label{h_c}
\end{equation}
with $\mb{S}_{j,\alpha}$ being the spin-1 operator at site $j$ of chain $\alpha$. Here $J>0$ is  such that the first term represents an antiferromagnetic exchange coupling, while the second term corresponds to a biquadratic interaction  tuned to a critical point  at which the model is exactly solvable by the Bethe ansatz \cite{TakhtajanPLA1982, BabujianPLA1982}. The chains are coupled at their end sites $j=1$ by the boundary interactions \begin{equation}
   H_{\mc{B}}=J_{\chi}\mathbf{S}_{1,1}\cdot\left(\mathbf{S}_{1,2}\times\mathbf{S}_{1,3}\right)+J^{\prime}\sum_{\alpha=1}^{3}\mathbf{S}_{1,\alpha}\cdot\mathbf{S}_{1,\alpha+1}\label{h_b}.
\end{equation}
These interactions  preserve SU(2) symmetry in addition to a $\mathbb{Z}_3$ symmetry under a cyclic permutation of the chain index $\alpha$, i.e., $\alpha\mapsto \alpha+1$ $(\text{mod } 3)$. Note that the $J_\chi$ interaction involves the scalar spin chirality for the boundary spins. This three-spin interaction  breaks   reflection ($\mc P$)    and time reversal ($\mc T$) symmetries,   \be
\mathcal{P}:\,\alpha\mapsto -\alpha \, (\text{mod } 3),\qquad  \mathcal{T}:\,\mathbf{S}_{j,\alpha}\rightarrow -\mathbf{S}_{j,\alpha}, \label{PT}\ee
but preserves the product $\mathcal{PT}$.

The low-energy excitations of each spin chain with length $L\gg1$ are described by an SU$(2)_2$ WZNW model  \cite{Gogolin2004, DiFrancesco1997}.  Before imposing boundary conditions, we can write the effective Hamiltonian in the Sugawara  form
\begin{equation}
H_{c} = \sum_{\alpha=1}^3\frac{\pi v }{2}\int_0^{L} dx\,(\mathbf{{J}}_\alpha^2+\bar{\mathbf{{J}}}_\alpha^2).
\label{wzweffect}
\end{equation}
Here $v\sim J$ is the spin velocity and $\mb{J}_\alpha$ and $\bar{\mb{J}}_\alpha$ are   left- and right-moving  currents, respectively, that obey the SU$(2)_2$ Kac-Moody algebra.  

 All local operators in the SU$(2)_2$ WZNW model can  be represented in terms of three critical Ising models \cite{Tsvelik1990,Allen2000}. In particular, the   currents are written as bilinears of chiral Majorana fermions $\xi^a_\alpha$ and $\bar \xi^a_\alpha$:
\begin{align}
{{J}}^{a}_{\alpha}(x)&=-\frac{i}{2}\epsilon^{abc}\xi_{\alpha}^{b}(x)\xi_{\alpha}^{c}(x),\nonumber\\ \Bar{{J}}^{a}_{\alpha}(x)&=-\frac{i}{2}\epsilon^{abc}\Bar{\xi}_{\alpha}^{b}(x)\Bar{\xi}_{\alpha}^{c}(x),
    \label{current-majorana}
\end{align}
where $a,b,c\in\{x,y,z\}\equiv\{1,2,3\}$ and $\epsilon^{abc}$ is the Levi-Civita symbol.  For each chain, the chiral Majorana fermions transform as a vector  $\xi_\alpha=(\xi^1_\alpha,\xi^2_\alpha,\xi^3_\alpha)^t$ under spin rotations. Combining  right and left movers, we define the components of the spin-1 primary  matrix  field\be
\Phi^{(1)}_{\alpha,ab} (x)= i\xi^a_\alpha(x)\,\bar{\xi}^b_\alpha(x),\label{phi-1}
\ee
which has scaling dimension 1.  The  diagonal elements  of the spin-1 field  can be identified with the energy operators in the Ising CFT, $\varepsilon^a_\alpha = i\xi^a_\alpha\,\bar{\xi}^a_\alpha$. In this representation, the Hamiltonian in Eq. (\ref{wzweffect}) becomes\be
H_{c}=\sum_{\alpha,a} \int_{0}^{L}dx\,\frac{iv}{2} (\xi^{a}_{\alpha}\partial_{x}\xi^{a}_{\alpha}-\bar\xi^{a}_{\alpha}\partial_{x}\bar\xi^{a}_{\alpha}).
\ee
The critical point is perturbed by one relevant bulk operator, which can be written as a mass term for the Majorana fermions:\be
\delta H_m=im\sum_{a,\alpha}\int_0^Ldx\, \xi^a_\alpha\bar \xi^a_\alpha. \label{massterm}
\ee
Tuning the strength of the biquadratic interaction in Eq. (\ref{h_c}) is equivalent to setting $m=0$ in the effective field theory. The Haldane phase and the dimerized phase correspond to $m>0$ and $m<0$, respectively; see Refs. \cite{Allen2000,XavierPRB2022}.

The theory  also contains a spin-$1/2$ primary matrix  field $\Phi^{(\frac12)}$ with scaling dimension $3/8$. The  components of $\Phi^{(\frac12)}$ can be expressed using  the order ($\sigma$) and disorder ($\mu$) Ising operators: \begin{align}
    \text{tr} \left[\Phi^{(\frac12)}_{\alpha}(x) \right]&\sim\sigma^{1}_{\alpha}\sigma^{2}_{\alpha}\sigma^{3}_{\alpha},\label{primaryrep1}\\ 
    \text{tr} \left[\tau^{a}\Phi^{(\frac12)}_{\alpha}(x) \right]&\sim\sigma^{a}_{\alpha}\mu^{a+1}_{\alpha}\mu^{a+2}_{\alpha}\,,
    \label{primaryrep2}
\end{align}
where $\tau^a$ are Pauli matrices.  For each chain,  these operators satisfy the relations
\begin{align}
    \sigma_\alpha^{a}(x)\mu_\alpha^{a}(y)&=\mu_\alpha^{a}(y)\sigma_\alpha^{a}(x)\text{sgn}(x-y),\\
    \sigma_\alpha^{a}(x)\xi_\alpha^{a}(y)&=\xi_\alpha^{a}(y)\sigma_\alpha^{a}(x)\text{sgn}(x-y),
    \label{order-majorana}\\
    \mu_\alpha^{a}(x)\xi_\alpha^{a}(y)&=-\xi_\alpha^{a}(y)\mu_\alpha^{a}(x)\text{sgn}(x-y).\label{disorder-majorana}
\end{align}
The spin-$1/2$ field appears, for instance, in     the staggered part of the spin operator in the continuum:\be
    \mathsf{\mathbf{S}}_{j,\alpha}\sim \mathsf{\mathbf{J}}_{\alpha}(x)+\Bar{\mathsf{\mathbf{J}}}_{\alpha}(x)+(-1)^{j}\mathsf{\mathbf{n}}_{\alpha}(x)\label{spinop}
\ee
where $\mathsf{\mathbf{n}}_{\alpha}(x)=\mathcal{A} \, \text{tr} [\mathsf{\boldsymbol{\tau}}\Phi^{(\frac12)}_{\alpha}(x) ]$  with a nonuniversal prefactor  $\mathcal{A}$. Besides the specific model  in Eq. (\ref{h_c}), the SU(2)$_2$ WZNW universality class can also be realized at the dimerization transition of  antiferromagnetic Heisenberg chains with three-site interactions \cite{Michaud2012,ChepigaAffleckPRB2016}.

\begin{figure}
    \centering
    \includegraphics[width=0.9\columnwidth]{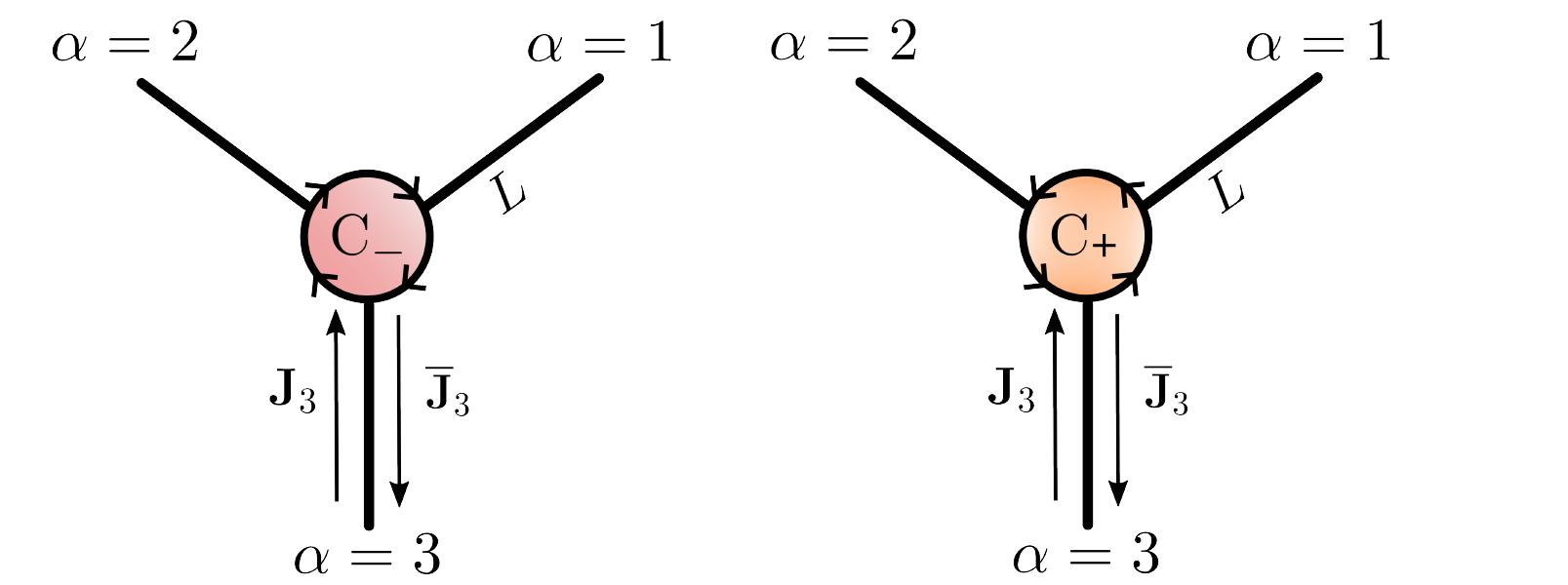}
    \caption{Schematic  representation of a junction of spin-1 chains.   In the continuum limit, the low-energy modes in each chain $\alpha$ are described by SU(2)$_2$  currents $\mathsf{\mathbf{J}}_{\alpha}(x)$ and $\Bar{\mathsf{\mathbf{J}}}_{\alpha}(x)$. At the chiral fixed points, denoted by C$_-$ and C$_+$, the currents are perfectly transmitted in the directions represented by the   clockwise or counterclockwise loops.}
    \label{fig1}
\end{figure}

The microscopic  interactions in Eq. (\ref{h_b}) can be tuned to control the boundary conditions for the low-energy modes at the junction.  Two chiral fixed points with opposite chirality, denoted as C$_+$ and C$_-$,  occur at intermediate  values of $J_\chi$ and $J'$ \cite{XavierPRB2022}. They are characterized by the boundary conditions\be
\text{C}_\pm:\;\mb J_\alpha(0)=\bar{\mb J}_{\alpha\pm 1}(0)
\ee
At the $\text{C}_{\pm}$ points, the Y junction behaves as an ideal spin circulator \cite{BuccheriPRB2018}, in which incoming spin currents are completely transmitted from one chain to the next  in rotation, either clockwise or counterclockwise; see Fig. \ref{fig1}. The direction of circulation is controlled by the sign of $J_\chi$ at the corresponding chiral fixed point. The chiral boundary conditions can be implemented in terms of Majorana fermions as\be
\text{C}_\pm:\;  \xi_\alpha^a(0)= p_\alpha \bar{\xi}^a_{\alpha\pm1}(0) , 
\ee
where $p_\alpha\in \{+1,-1\}$ can be chosen arbitrarily, manifesting a $\mathbb Z_2$ gauge freedom in the fermionic representation of Eq. (\ref{current-majorana}). Fixing $p_\alpha=+1$, we can glue the fermionic modes in different chains as\be
\text{C}_\pm:\;   {\xi}_\alpha^a(x)= \bar{\xi}^a_{\alpha\pm1}(-x). 
\ee
Thus, we regard the left-moving Majorana fermions  as the analytic continuation of the right-moving ones to the domain $x<0$. With this convention, the effective Hamiltonian for a single junction tuned to a chiral fixed point can be cast  in the form 
\begin{align}
	H_{Y}^{\rm CFP}=-\frac{iv}{2}\sum_{\alpha,a} \int_{-L}^{L}dx\, \bar\xi^{a}_{\alpha}\partial_{x}\bar\xi^{a}_{\alpha}.
	\label{h_c-Majorana}
\end{align}

The chiral-fixed-point Hamiltonian is perturbed by one relevant and one marginal boundary operator. These operators are written in terms of the trace of the primary fields at $x=0$:\bea
\delta H&=&\gamma \sum_\alpha \text{tr}[\Phi_\alpha^{(\frac12)}(0)]+\lambda \sum_\alpha \text{tr}[\Phi_\alpha^{(1)}(0)].\label{deltaH}
\eea
The coupling constants $\gamma$ and $\lambda$ vanish when the microscopic parameters $J_\chi$ and $J'$ are fine tuned to one of the chiral fixed points. 
The relevant $\gamma$ interaction drives the system towards low-energy fixed points with vanishing spin conductance \cite{XavierPRB2022}. The marginal $\lambda$ interaction can be written   in terms of   Majorana fermions and  corresponds to a backscattering process at the junction. These perturbations render  the chiral fixed points   unstable    in the limit $L\to\infty$. 
 However, if the crossover to stable fixed points happens to be slow, as verified numerically for the junction of spin-1/2 chains \cite{BuccheriPRB2018,BuccheriNPB2019}, the chiral fixed point can still govern the physical properties of a junction with finite but very long  chains  over a wide range of parameters.  Moreover, we can cut off the infrared divergence  of the relevant perturbation by keeping the chain length finite and imposing boundary conditions at $x=L$ that correspond to constructing a 2D network, as we will discuss in the following.

\section{ Network model with staggered chirality}\label{section-3}

\begin{figure}
    \centering
    \includegraphics[scale=0.45]{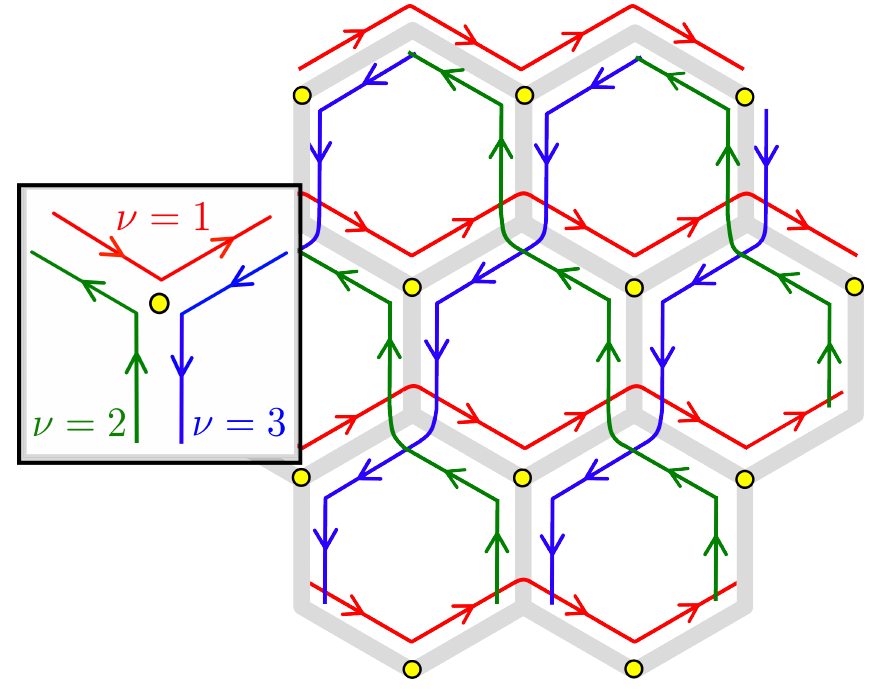}
    \caption{Network of junctions at the chiral fixed point. Inside the box, we represent the unit cell specifying the direction of propagation of the three Majoranas chiral modes. The origin of the   unit cell is represented by the yellow dots, where we have $s=0$.  }
    \label{fig2}
\end{figure}

Consider a honeycomb network constructed  by putting together Y junctions of spin-1 chains. To obtain a translationally invariant system, we impose chiral boundary conditions with the same    chirality, say the C$_-$ fixed point, on all the junctions marked by yellow dots in Fig. \ref{fig2}. These positions  correspond to the $x=0$ end of the spin chains. The choice of the boundary conditions at  $x=L$  is crucial. If we impose the same chirality   as at $x=0$, we obtain the  non-Abelian  CSL discussed in Ref. \cite{XavierScipost2023}. By contrast, here we assemble a  network     with staggered chirality in order to obtain a gapless phase. This can be accomplished by tuning  the  interactions in  Eq. (\ref{h_b}) among the chains that meet at   $x=L$   to the C$_+$ fixed point. We can write the boundary conditions as
\begin{align}
    \Bar{{J}}^{a}_{\alpha,\mathbf{R}}\left(0\right)&={J}^{a}_{\alpha+1,\mathbf{R}}\left(0\right),\nonumber\\
    \Bar{{J}}^{a}_{\alpha,\mathbf{R}}\left(L\right)&={J}^{a}_{\alpha-1,\mathbf{R}-\boldsymbol{\delta}_{\alpha-1}}\left(L\right)
    \label{current-cbd2},
\end{align}
where the lattice vector $\mb{R}$ specifies the positions represented as yellow dots in Fig. \ref{fig2}, which form a triangular lattice, and  $\boldsymbol{\delta}_{\alpha}$ are the next-nearest-neighbor vectors 
\begin{align}
    \boldsymbol{\delta}_1&=\sqrt{3}L\left(1,0\right),\nonumber\\
    \boldsymbol{\delta}_2&=\sqrt{3}L\left(-1/2,\sqrt{3}/2\right),\nonumber\\
    \boldsymbol{\delta}_3&=\sqrt{3}L\left(-1/2,-\sqrt{3}/2\right).
\end{align}
In terms of the Majorana fermions, the chiral boundary conditions in Eq.  \eqref{current-cbd2} can be expressed as
\begin{align}
    \Bar{\xi}^a_{\alpha,\mathbf{R}}\left(0\right)&=p_{\alpha,\mb R}\,\xi^a_{\alpha+1,\mathbf{R}}\left(0\right),\nonumber\\
    \Bar{\xi}^a_{\alpha,\mathbf{R}}\left(L\right)&=p_{\alpha,\mb R}\,\xi^a_{\alpha-1,\mathbf{R}-\boldsymbol{\delta}_{\alpha-1}}\left(L\right)
    \label{cbd2},
\end{align}
with $p_{\alpha,\mb R}\in \{+1,-1\}$. Here we will fix   a uniform sign $p_{\alpha,\mb R}=+1$ $\forall\,\alpha, \mb R$, but will reexamine this choice later when we discuss the $\mathbb Z_2$ gauge structure of the resulting 2D phase.

With this choice of boundary conditions, the   network model describes  three sets of decoupled chiral 1D modes running along the  zigzags of the honeycomb lattice. The three directions of propagation are schematically represented by   red, green and blue lines  in Fig. \ref{fig2}. These modes are  related to each other   by the $\mathbb Z_3$ symmetry that combines a C$_3$ lattice rotation with  $\alpha\mapsto \alpha+1$. To specify positions in this network, we use the   coordinates $(\nu, \mb R, s)$, where  $\nu=1,\,2,\,3$ labels  the direction of propagation (red, green and blue, respectively), $\mb R$ is the center of the  unit cell  shown in the inset in Fig. \ref{fig2}, and $s\in [-L,L]$ is a continuous   coordinate within the unit cell, with $s=0$ at  the center of the unit cell and $s=\pm L$ at the far ends of the  chains. In this notation, the Majorana fields obey the relation   \be
  \bar  \xi^{a}_{\nu,\mathbf{R}-\boldsymbol{\delta}_{\nu}}(s+2L)= \bar\xi^{a}_{\nu,\mathbf{R}}(s).
\ee
The chiral fixed point Hamiltonian can be written as
\begin{align}
    H_{0}&=\frac{v}{2}\sum_{a,\nu}\sum_{\mathbf{R}}\int_{-L}^{L}ds\,\bar\xi^{a}_{\nu,\mathbf{R}}\left(s\right)(-i\partial_{s})\bar\xi^{a}_{\nu,\mathbf{R}}\left(s\right)
    \label{H0-Majoranas}.
\end{align}

As a consequence of the SU(2) symmetry, the Majorana fermions are degenerate with respect to the spin index $a$. It is convenient to single out the $z$ spin direction and define a complex fermion by combining two Majoranas:
\begin{align}
    \bar \psi_{\nu,\mathbf{R}}(s)&=\frac{1}{\sqrt{2}}\left[\bar \xi^{x}_{\nu,\mathbf{R}}(s)+i\bar\xi^{y}_{\nu,\mathbf{R}}(s)\right],
    \label{complex-fermions}
\end{align}
with $\{\bar\psi^{\phantom\dagger}_{\nu,\mathbf{R}}(s),\bar\psi^\dagger_{\nu',\mathbf{R}^{\prime}}(s^{\prime})\}=\delta_{\nu\nu'}\delta_{\mathbf{R},\mathbf{R}^{\prime}}\delta(s-s^{\prime})$. We can then write $H_0=H_0^{xy}+H_0^z$, with \bea
     H_{0}^{xy}&=&v \sum_{\nu,\mathbf{R}}\int_{-L}^{L}ds\, \bar\psi^{\dagger}_{\nu,\mathbf{R}}\left(s\right)(-i\partial_{s})\bar\psi_{\nu,\mathbf{R}}\left(s\right),\label{H0xy}\\
     H_0^z&=&\frac{v}{2}\sum_{\nu,\mb R}\int_{-L}^{L}ds \,\bar\xi^{z}_{\nu,\mathbf{R}}\left(s\right)(-i\partial_{s})\bar\xi^{z}_{\nu,\mathbf{R}}\left(s\right).
     \label{H0-fermions-R}
\eea
Hereafter we focus on the spectrum of the complex fermion, but we should keep in mind that the theory also contains the Majorana fermion $\bar \xi^z$ with the same dispersion relation but half the number of modes.

The quadratic Hamiltonian in Eq. (\ref{H0xy}) can be diagonalized straightforwardly. We use the mode expansion  
\begin{align}
     \bar \psi_{\nu,\mathbf{R}}\left(s\right)&=\frac{1}{\sqrt{2\mathcal{N}L}}\sum_{\mb{k}\in \text{BZ}}\sum_{n\in \mathbb{Z}}e^{i\left[\mb{k}\cdot\mathbf{R}+Q_{\nu,n}\left(\mb{k}\right)s\right]}\bar\psi_{\nu,n}\left(\mb{k}\right),
     \label{FT}
\end{align}
where   $\mathcal{N}$ is the number of unit cells and BZ stands for the first Brillouin zone. The auxiliary function $Q_{\nu,n}\left(\mb{k}\right)$ is defined as
\begin{align}
    Q_{\nu,n}\left(\mb{k}\right)=\frac{\mb{k}\cdot\boldsymbol{\delta}_\nu +2\pi n}{2L},\quad n\in\mathbb{Z}\,,
    \label{Q-function}
\end{align}
and obeys $Q_{\nu,\,-n}(-\mb k)=-Q_{\nu,\,n}(\mb k)$. This function  is important to ensure the relation $\bar\psi_{\nu,\,\mb R-\boldsymbol{\delta}_\nu}(s+2L)=\bar\psi_{\nu,\,\mb R}(s)$. The  Hamiltonian  can be written in momentum space as
\begin{align}
    H^{xy}_{0}=\sum_{\nu=1}^{3}\sum_{\mb{k}\in \text{BZ}}\sum_{n\in \mathbb{Z}}\mc E_{\nu,n}(\mb{k}) \bar\psi^{\dagger}_{\nu,n}(\mb{k})\bar\psi_{\nu,n}(\mb{k}) ,
    \label{H0-fermions-k}
\end{align}
where \be
\mc E_{\nu,n}(\mb{k})=vQ_{\nu,n}(\mb{k})\ee
is the  fermion dispersion relation with $n\in\mathbb{Z}$ denoting a band index.  It follows  from the definition of $Q_{\nu,n}(\mb{k})$ that  shifting the momentum by a reciprocal lattice vector,  $\mb k\mapsto \mb k+\mb G$ with $\mb{G}\cdot\boldsymbol{\delta}_{\nu}=2\pi\ell$ and $\ell\in \mathbb Z$, corresponds to shifting the band index $n\mapsto n+\ell$. Due to the continuum of states inside the unit cell, the spectrum of $H^{xy}_{0}$ exhibits an infinite number of positive- and negative-energy bands.  The ground state is  a Fermi sea in which  all negative-energy state states are occupied.  We stress that this field theory approach is aimed at describing the low-energy properties of the network. The low-energy bands correspond to $n=0$ and have the  dispersion relation\be
\mc E_{\nu,0}(\mb{k})=\frac{\sqrt3}2 v \mb{k}\cdot \hat{\mb e}_\nu,
\ee
where $\hat{\mb e}_\nu=\boldsymbol \delta_\nu/(\sqrt3L)$ are unit vectors. There are three low-energy bands that disperse along  the directions of propagation of the chiral 1D modes.  The spectrum is gapless along three intersecting straight lines in reciprocal space  given by $\mb{k}\cdot \hat{\mb e}_\nu=0$.

 \begin{figure*}
    \centering
    \includegraphics[width=0.95\textwidth]{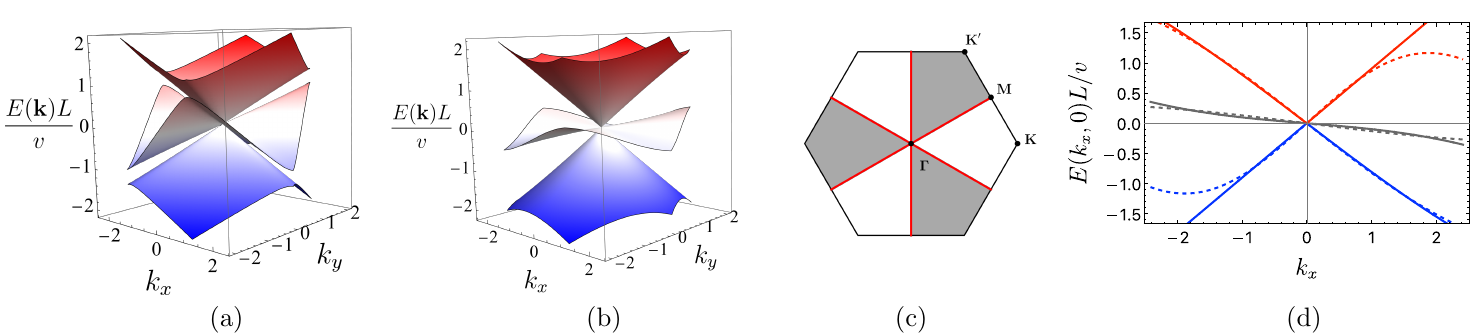}
    \caption{Dispersion relation  of fermionic excitations in the network model, see Eq. (\ref{Hxyeff}), for two values of  the coupling constant of the backscattering operator:  (a) $\lambda/v=0.2$; (b) $\lambda/v=0.8$. Panel (c) shows the Fermi surface of the middle band (red lines), with occupied states corresponding to the shaded regions. The  primitive  reciprocal lattice vectors are   $\mb{b}_{1}=\left(\frac{2\pi}{\sqrt{3}L},\frac{2\pi}{3L}\right)$ and  $\mb{b}_{2}=\left(0,-\frac{4\pi}{3L}\right)$.   In panel (d), the dispersion for  $\lambda/v=0.8$ (solid lines) is compared with the tight-binding model of Eq. (\ref{Htb}) with $t_1=0.78v/L$ and $t_3=1.09v/L$ (dashed lines). Here we set $E_{\rm max}=v/L$ as the energy cutoff for the fitted data.}
    \label{fig3}
\end{figure*}

The solution of the chiral-fixed-point Hamiltonian in terms of decoupled 1D modes has  a direct impact on  the spin correlation. Let $\mb S_{j,\alpha,\mb R}$ denote the spin operator at site $j$ of chain $\alpha$ of  the junction centered at $\mb R$. Given the chain index $\alpha$, the two chiral modes that run through this chain propagate along the directions labeled as $\nu=\alpha$ and $\nu=\alpha-1$; see Figs. \ref{fig1} and \ref{fig2}. At the chiral fixed point, two spins separated by a distance $r> L$ are correlated only if there is  a chiral mode that connects them. This condition requires that the second point be located at a unit cell given by $\mb R'=\mb R+m\boldsymbol\delta_{\alpha}$ or $\mb R'=\mb R+m\boldsymbol\delta_{\alpha-1}$ with $m\in\mathbb Z$. We can then compute the correlation using the representation of the spin operator in Eq. (\ref{spinop}) and   the operator product expansion (OPE) of the SU(2)$_2$ currents  \cite{Gogolin2004, DiFrancesco1997}\be
\Bar{{J}}^{a}(\Bar{z})\Bar{{J}}^{b}(\Bar{z}^{\prime})\sim\frac{1}{4\pi^{2}}\frac{\delta^{ab}}{(\Bar{z}-\Bar{z}^{\prime})^{2}}+i\epsilon^{abc}\frac{\Bar{{J}}^{c}_{\alpha}(\Bar{z}^{\prime})}{2\pi(\Bar{z}-\Bar{z}^{\prime})},\ee
where $\bar z=v\tau-ix$  is the anti-holomorphic  coordinate in Euclidean spacetime. We obtain  \be
    \langle {S}^z_{j,\alpha,\mb R}  {S}^z_{j',\alpha^{\prime},\mb R'}\rangle \sim-\frac{\delta_{\mb R',\mb R+m\boldsymbol\delta_{\alpha}}+\delta_{\mb R',\mb R+m\boldsymbol\delta_{\alpha-1}}}{4\pi^2(s'-s+2mL)^2},\label{correlation}
\ee
where on the right-hand side   $s,s'\in[-L,L]$ are coordinates within   the   unit cells corresponding to the sites $(j,\alpha,\mb R)$ and $(j',\alpha',\mb R')$, respectively, for  the chiral model that connects these two points. Thus, the correlation is spatially anisotropic and decays as a power law with the distance  along the special directions set by the vectors $\boldsymbol\delta_\alpha$. Remarkably, the staggered part of the correlation vanishes  because the spin-1/2 primary field acts nontrivially on both chiral sectors of a given chain, and two points  separated by a distance $r>L$ cannot share both chiral modes. The same behavior was obtained  within a different coupled-wire construction for a model with staggered chirality on the extended kagome lattice \cite{PereiraScipost2018}.

\section{Perturbations to the chiral fixed point \label{section 4}}

So far we have explored the physics of the network model when the microscopic interactions   are tuned to the chiral fixed points. An immediate question concerns what happens   in the presence of perturbations associated with deviations from the chiral fixed point. As discussed in Sec. \ref{Yjunctionsec}, for a single junction there are   two non-irrelevant boundary interaction terms given by Eq. (\ref{deltaH}). We now analyze the effects of these perturbations on the excitation spectrum of the network. We start with the marginal operator, which involves the spin-1 field and can be treated exactly, and then proceed to the analysis of the operator that involves the spin-1/2 field.  We also consider the effect of the relevant mass term in Eq. (\ref{massterm}). We will see that the chiral fixed point is unstable against the operators that are relevant at the 1D fixed point, but a finite strength of the marginal coupling can stabilize a 2D gapless phase.

\subsection{Spin-1 boundary perturbation: backscattering of Majorana fermions\label{spin1section}}

Let us consider the second term in Eq.  (\ref{deltaH}). Using the fermionic representation and imposing either C$_+$ or C$_-$ chiral boundary conditions, we can write this term as \be
\delta H^{(\text{C}_{\pm})}_\lambda=i\lambda \sum_{a,\alpha}  \bar\xi^a_{\alpha\pm 1}(0) \bar\xi^a_{\alpha}(0). 
\ee
This operator corresponds to a  backscattering process that hybridizes the chiral modes at $x=0$. For a single junction,  this marginal perturbation defines a critical  line in the boundary phase diagram where the spin conductance tensor can be calculated exactly \cite{XavierPRB2022}. When transported to the network with staggered chirality, the perturbation becomes 
\begin{align}
    H_{\lambda}= -i\lambda\sum_{\eta=0,1}\sum_{a,\nu}\sum_{\mathbf{R}}(-1)^{\eta}\bar \xi^{a}_{\nu+1,\mathbf{R}} (\eta L )\bar\xi^{a}_{\nu,\mathbf{R}-\eta\boldsymbol{\delta}_{\nu}} (\eta L ).
    \label{Hbs-fermions-R}
\end{align}

Importantly,  the sign  of the backscattering amplitude alternates between  $s=0$ and $s=L$. This property  is related to a mirror symmetry of the network model with   staggered chirality. Consider the reflection with respect to a horizontal line that runs through the center of   a hexagon in Fig. \ref{fig2}. We define the vectors $\mb w_1=L(\frac{\sqrt3}2,\frac12)$, $\mb w_2=L(\frac{\sqrt3}2,-\frac12)$ and $\mb w_3=L(0,-1)$, such that $\boldsymbol\delta_\alpha=\mb w_\alpha-\mb w_{\alpha+1}$.  The mirror symmetry is implemented as  \begin{align}
 {\bar\xi}_{1,\mathbf{R}}(s) &\mapsto \begin{cases}
\bar\xi_{1,\tilde{\mathbf{R}}-\mb{w}_{2}}(s-L) & \mbox{if $s\in(0,L)$},\\
\bar\xi_{1,\tilde{\mathbf{R}}-\mb{w}_{1}}(s+L) & \mbox{if $s\in(-L,0)$},\end{cases}\nonumber\\
 {\bar\xi}_{2,\mathbf{R}}(s) &\mapsto \begin{cases}
\bar\xi_{3,\Tilde{\mathbf{R}}-\mb{w}_{1}}(s-L) & \mbox{if $s\in(0,L)$},\\
\bar\xi_{3,\Tilde{\mathbf{R}}-\mb{w}_{3}}(s+L) & \mbox{if $s\in(-L,0)$},\end{cases} \nonumber\\
 {\bar\xi}_{3,\mathbf{R}}(s) &\mapsto \begin{cases}
	\bar\xi_{2,\Tilde{\mathbf{R}}-\mb{w}_{3}}(s-L) & \mbox{if $s\in(0,L)$},\\
	\bar\xi_{2,\Tilde{\mathbf{R}}-\mb{w}_{2}}(s+L) & \mbox{if $s\in(-L,0)$},\end{cases}
\label{AM3}
\end{align}
where $\Tilde{\mathbf{R}}$ stands for the unit cell position   after the  reflection. It is straightforward to check that the  operator in Eq. \eqref{Hbs-fermions-R} is invariant under this transformation, but only if the relative minus sign is properly taken into account. This   symmetry should not be confused with $\mc P$ defined in Eq. (\ref{PT}), which refers to a reflection about a vertical line  through the center of a hexagon. The latter  inverts the direction of propagation of all chiral modes, but it can    be combined with time reversal to yield the $\mc P\mc T$ symmetry of the network model.

The effective Hamiltonian including the marginal perturbation is quadratic in the Majorana fermions and can be diagonalized exactly. Once again, we focus on the contribution from the complex fermion in Eq. (\ref{complex-fermions}) and use the mode expansion in Eq. (\ref{FT}). The leading effect of the backscattering term is to generate avoided level crossings of nearly degenerate states, opening gaps between bands in close analogy with the band structure of electrons in a weak periodic potential \cite{ashcroftsolid}.   Truncating the spectrum to keep only the low-energy bands with index $n=0$, we can write  \be
H^{xy}=H_0^{xy}+H_\lambda^{xy}=\sum_{\mb k} \Psi^\dagger (\mb k)\mc H_{\rm eff}(\mb k)\Psi(\mb k),\label{Hxyeff}
\ee
where $\Psi(\mb{k})=\left(\bar\psi_{1,0}(\mb{k}),\bar \psi_{2,0}(\mb{k}),\bar\psi_{3,0}(\mb{k})\right)^{\text{T}}$. The Bloch Hamiltonian reads \be
\mathcal{H}_{\text{eff}}(\mb{k})=\left(\begin{array}{ccc}
\mc E_{1,0}(\mb k) & \Lambda_{12}(\mb{k}) & \Lambda_{31}^{*}(\mb{k})\\
\Lambda_{12}^{*}(\mb{k}) & \mc E_{2,0}(\mb k) & \Lambda_{23}(\mb{k})\\
\Lambda_{31}(\mb{k}) & \Lambda_{23}^{*}(\mb{k}) & \mc E_{3,0}(\mb k)\end{array}\right),\label{Bloch}
\ee
where the off-diagonal elements  are given by \be
\Lambda_{\nu,\nu+1}(\mb k)= \frac{i\lambda}{2L}\left(1-e^{-i\mb{k}\cdot\boldsymbol{\delta}_{\nu-1}/2}\right).\label{Lambda}
\ee
Note that $\Lambda_{\nu,\nu+1}(\mb k)$ vanishes for $\mb k\to0$ as a result of the negative interference between the scattering processes at $x=0$ and $x=L$.

Diagonalizing $\mathcal{H}_{\text{eff}}(\mb{k})$, we obtain closed-form but lengthy expression for the dispersion relations  of the fermionic bands, denroted as $E_r(\mb k)$ with $r=1,2,3$. The result is shown in Fig. \ref{fig3}.  The bands have the property   $E_{r}(-\mb k)=-E_{4-r}(\mb k)$. In particular, $E_2(\mb k)$ transforms into itself under  $\mb k\mapsto -\mb k$.  The gapless lines of this middle band  are  determined by the zeros of the determinant of $\mathcal{H}_{\text{eff}}(\mb{k})$.  Using $\sum_\nu\boldsymbol\delta_\nu=0$, it is easy to show that the gapless lines occur at $\mb k\cdot \hat{\mb e}_\nu=0$, which is the same Fermi surface that  we obtained   for the chiral fixed point.  The bandwidth of $E_2(\mb k)$ decreases as   we increase the ratio $\lambda/v$. In addition to the Fermi surface of the middle band, the lower and upper bands touch zero energy with a Dirac cone at the $\Gamma$ point.  This band touching can be understood by noting that $\mathcal{H}_{\text{eff}}(\mb{k})$ vanishes for $\mb k=0$.

A qualitatively similar spectrum has been obtained    in  parton mean-field descriptions of gapless CSLs on the kagome lattice \cite{PereiraScipost2018, BauerPRB2019, OlivieroSciPost2022}. Note, however, that the result obtained here does not rely on mean-field approximations because the fractionalization into Majorana fermions is   established within the building blocks, namely the critical spin-1 chains. To make the connection with parton mean-field theory more explicit, we can fit the low-energy spectrum of the effective Hamiltonian to a tight-binding model. This approximation is better justified when  we increase  the coupling constant of the marginal operator so that  the band splitting, determined by the off-diagonal matrix elements of order $|\lambda|/L$ in Eq. (\ref{Lambda}), becomes comparable to the bandwidth $W\sim v/L$ of the unperturbed model.

 \begin{figure}
    \centering
    \includegraphics[width=.9\columnwidth]{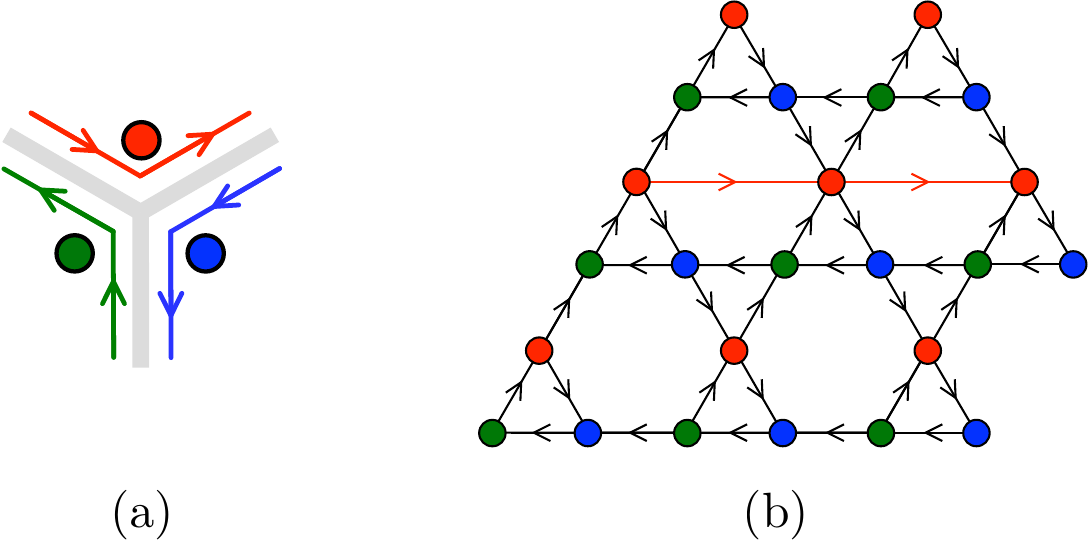}
    \caption{Mapping   to a tight-binding model. (a) Each  chiral mode ($\nu=1,2,3$ for red, green, and blue, respectively) of the  $n=0$ bands   is mapped onto a single site. (b) The propagation of the chiral modes can be represented on an extended kagome lattice. The red line indicates the direction of free propagation of the  $\nu=1$ mode; cf. Fig. \ref{fig2}. Similarly, the modes $\nu=2,3$ run along the other diagonals of the hexagons.  The arrows in the black lines represent the orientation of the links in the backscattering term; see Eq. (\ref{Htb}). }
    \label{fig4}
\end{figure}

The mapping of the network model to a tight-binding model can be visualized as shown  in Fig. \ref{fig4}. We represent the   chiral modes in the low-energy bands as three sites forming a triangle and assign  three Wannier states $|\nu,\mb R\rangle$ to each unit cell. Putting  the unit cells together, we naturally obtain a  kagome lattice. The free propagation of the chiral modes at the chiral fixed point, see Fig. \ref{fig2}, corresponds to hopping along the diagonals of the hexagons, which connect sites that belong to the same sublattice. The backscattering processes at the junctions  are mapped onto hoppings between nearest-neighbor sites, which belong to  different sublattices and form the triangles of the kagome lattice. The minimal tight-binding model compatible with the symmetries of the model is 
\bea
H^{xy}_{\rm tb}&=&\frac{ i t_1}2\sum_{\nu,\mb R}\left[ (f^\dagger_{\nu+1,\mb R- \mb a_{\nu-1}}-f^\dagger_{\nu+1,\mb R})f^{\phantom\dagger}_{\nu,\mb R}-\text{h.c.}\right]\nonumber\\
&&+\frac{i t_3}2\sum_{\nu,\mb R}\left(f^\dagger_{\nu,\mb R+ \mb a_{\nu}} f^{\phantom\dagger}_{\nu,\mb R}-\text{h.c.}\right),\label{Htb}
\eea
where $f_{\nu,\mb R}$ annihilates  a fermion in the state $|\nu,\mb R\rangle$, $t_1$ and $t_3$ are the hopping parameters of order $v/L$, and we set $\mb a_\nu=\boldsymbol\delta_{\nu}/2$ to match the off-diagonal matrix elements in Eq. (\ref{Lambda}). As a result, the network model describes a staggered-flux ansatz \cite{BieriPRB2016}   with first- and third-neighbor links on a kagome lattice. Note that a second-neighbor imaginary hopping $it_2$ is forbidden because it would break the reflection symmetry of the network model. We can fit the hopping parameters to reproduce the low-energy spectrum by minimizing the mean square deviation for the dispersion relation in the range $|E_r(\mb k)|<E_{\rm max}$ with $E_{\rm max}\sim v/L$; see Fig. \ref{fig3}(d). Importantly, the ratio $t_1/t_3$ increases with $\lambda/v$ up to $\lambda/v\sim 1$, at which point the middle band becomes approximately flat.

The Majorana Fermi surface governs the low-energy properties of the CSL  on the 2D network.  For instance, the local dynamical spin correlation behaves as \begin{align}
   C(\omega)&=\int_{-\infty}^{\infty}dt\, e^{i\omega t}\langle S^{z}_{j,\alpha,\mb R}(t)S^{z}_{j,\alpha,\mb R}(0)\rangle\sim \omega,
\end{align}
analogous to  the local density of states of particle-hole excitations in a Fermi liquid. We can show that the  power-law decay of the equal-time spin correlation at  distances $r=|\mb r \cdot  \hat{\mb e}_\nu|\gg L$ predicted in Eq. (\ref{correlation}) remains valid   in the presence of the backscattering operator because  the  Fermi surface still has the form of straight lines \cite{PereiraScipost2018}. This is a slower decay than the $1/r^3$ behavior expected for a 2D Fermi surface with nonzero  curvature \cite{Motrunich2007}.  In addition,   quantum spin liquids with  a Fermi surface of fractional excitations are characterized by a logarithmic violation of the  area law for the entaglement entropy  \cite{WolfPRL2006, SwinglePRL2010}. In two dimensions, the entanglement entropy $S_E$ of a subsystem of linear size $\mc L$ in  the gapless CSL scales as $S_E\sim \mc L \ln \mc L$. In this network construction, the logarithmic correction of   the 2D phase is directly connected  to the entanglement entropy of the chiral 1D modes. The simple argument \cite{PereiraScipost2018} is that  the number of  1D modes that crosses the  boundary of the subsystem with linear size $\mc L\gg L$ is proportional to $\mc L$, and each mode contributes to the entanglement entropy with $S_E^{\rm 1D}\sim c \ln \mc L$, where $c$ is the central charge of the CFT \cite{CalabreseIOP2004}.

A remark about the nomenclature is in order. We  refer to this CSL state as ``Majorana Fermi surface'' rather than  ``spinon Fermi surface''   because we reserve the term ``spinon'' for  excitations that carry spin $1/2$, whereas the  gapless Majorana fermions  carry spin $1$ \cite{Tsvelik1992,Chen2012}. In the next section we will discuss the properties of the spin-1/2 excitations in the network model and show that they are related to gapped visons.

\subsection{Spin-1/2 boundary perturbation: visons}

We now turn to the perturbation described by the first term in Eq. (\ref{deltaH}). In the network model, we define \be
H_\gamma=\gamma \sum_{\eta=0,1}\sum_{\alpha}\sum_{\mb R} \sigma_{\alpha,\mb R}^1(\eta L)\sigma_{\alpha,\mb R}^2(\eta L)\sigma_{\alpha,\mb R}^3(\eta L),\label{Hgamma}
\ee
where we used the representation in Eq. (\ref{primaryrep1}). Unlike the other terms of the effective Hamiltonian   we have discussed so far, $H_\gamma$ cannot be written as a local operator in terms of Majorana fermions.  This operator is a relevant perturbation to  the chiral fixed point for a single junction in the limit $L\to \infty$ \cite{XavierPRB2022}. Moreover,  this perturbation destabilizes the chiral fixed point of the network with staggered chirality depicted in Fig. \ref{fig2}   because the latter is described in terms of decoupled 1D modes that extend to infinity.  However, in the presence of the marginal perturbation with $\lambda\sim v$, the fermionic excitations develop a 2D dispersion at low energies, see  Fig. \ref{fig3}, and the analysis based on the scaling dimension at the 1D fixed point no longer applies. On the other hand, we can investigate the effect of  $H_\gamma$    in this regime    using  the effective tight-binding model discussed in Sec. \ref{spin1section}.

First, we note that, if we come from the 1D limit and integrate out high-energy modes, the Majorana fermions become interacting in the presence of the $\gamma$ perturbation. To see this, we use the OPE of the order operator in the Ising CFT, represented by the fusion rule \cite{DiFrancesco1997}\be
\sigma\times \sigma =\mathbb{1}+\varepsilon,
\ee
where $\mathbb{1}$ denotes the identity operator. We   then treat $H_\lambda$ within second-order perturbation theory, applying the OPE to the $\sigma$ fields in the  independent spin sectors labeled by $a=1,2,3$. Integrating out high-energy modes, we generate a term linear in $\varepsilon_\alpha^a$, which amounts to a renormalization of the backscattering amplitude $\lambda$. In addition, we obtain a quadratic term in the energy operator  the form \be
H_{\rm int}=g\sum_{\mb R}  \sum_{\eta=0,1}\sum_{a,\alpha}\sum_{\mb R} \varepsilon_{\alpha,\mb R}^a(\eta L)\varepsilon_{\alpha,\mb R}^{a+1}(\eta L),\ee
which is a fermion-fermion  interaction  with coupling constant $g\sim \gamma^2$.  Were this the only effect of the $\gamma$ perturbation, we would expect the Majorana Fermi surface  in the network model to be completely robust  in the small-$\gamma$ limit. The reason is that chiral Fermi surfaces  are generically stable against weak short-range interactions \cite{Barkeshli2013,MotrunichPRB2011b,Chari2021}. In addition to the absence of nesting,  the conventional Cooper instability (with zero-momentum pairing) is  ruled out because states with opposite momentum are not degenerate when  time reversal and inversion symmetries are broken.

 However, as mentioned above, the chiral fixed point with $\lambda=0$ must be destabilized by an arbitrarily small  $\gamma$.  It is instructive to note  that in the network with uniform chirality the $\gamma$ perturbation has been shown to  create visons that  bind Majorana zero modes and behave as non-Abelian spinons \cite{XavierScipost2023}. Thus, we anticipate that this perturbation also creates visons in the gapless CSL, and the stability of the phase depends on whether visons become gapped excitations in the 2D regime   $\lambda\sim v$.

Beyond the approximation of  integrating out the $\sigma$ operators, the $\gamma$ interaction must be related to an emergent $\mathbb Z_2$ gauge field.  Let us  revisit the $\mathbb Z_2$ gauge degree of freedom   alluded to  in Eq.  (\ref{cbd2}). The variable $p_{\alpha,\mb R}\in\{+1,-1\}$ can be interpreted in terms of  the phase shift that the  Majorana fermion picks up when tunneling from one chain to the next according to the chiral boundary conditions.  At the chiral fixed point, a sign change in $p_{\alpha,\mb R}$ at any  junction can be gauged away since the chiral modes   are defined on open  lines that extend out to infinity. However, once we turn on the backscattering operator, the Majorana fermions   can   move around in closed paths and feel the physical effects of a gauge-invariant $\mathbb Z_2$ flux. 

When we set $p_{\alpha,\mb R}=+1$ $\forall \alpha,\mb R$ in the effective Hamiltonian, we assumed that the uniform gauge configuration describes  the sector of the Hilbert space that contains the ground state. For consistency, we must inquire about the energy cost of flipping the sign of $p_{\alpha,\mb R}$. The action of the operator in Eq. (\ref{Hgamma}) has precisely this effect because      $\sigma$ acts as a twist field that changes the boundary conditions for the Majorana fermions \cite{ChamonMudryPRB2019,Slagle2022,XavierScipost2023}. In fact, Eq. (\ref{order-majorana}) implies that the  fermions pick up a minus sign when they go around the point where   $\sigma$  is applied.

To estimate the energy of the gauge-field excitations,  we turn to the effective tight-binding model in Eq. (\ref{Htb}). We   implement the $\mathbb Z_2$ gauge degree of freedom by rewriting the  Hamiltonian as\be
H^{xy}_{\rm tb}=\frac{i}2\sum_{ij}u_{ij}t_{ij}f^\dagger_if^{\phantom\dagger}_j,\label{tbwithuij}
\ee
where $t_{ij}=t_1$ for first-neighbor links, $t_{ij}=t_3$ for third-neighbor links, and $t_{ij}=0$ otherwise. Here $u_{ij}\in\{+1,-1\}$ denotes an Ising link variable obeying the relation $u_{ji}=-u_{ij}$. In the proposed ground state, we fix the positive orientation of the links with nonzero $u_{ij}$ as represented  by    the arrows in Fig.   \ref{fig4}(b).  The gauge-invariant $\mathbb Z_2$ flux can be defined from the product of $u_{ij}$ around the plaquettes of the kagome lattice \cite{ChuaPRB2011,MotrunichPRB2011a,MotrunichPRB2011b}. The effect of the $\gamma$ perturbation in Eq. (\ref{Hgamma}) is mapped onto flipping the sign of the variable $u_{ij}$ that corresponds to a  path  on the network that contains the point where the order operator $\sigma$ is applied. As a result, the quantum fluctuations of $u_{ij}$ create  pairs of visons  on neighboring  plaquettes that share the link $(i,j)$. For $\gamma\neq0$, the $\mathbb Z_2$ gauge field becomes a dynamical degree of freedom. The situation here is analogous to the effect of  integrability-breaking interactions in the Kitaev spin liquid \cite{ZhangPRB2021,Joy2022}.

 \begin{figure}
    \centering
    \includegraphics[width=.95\columnwidth]{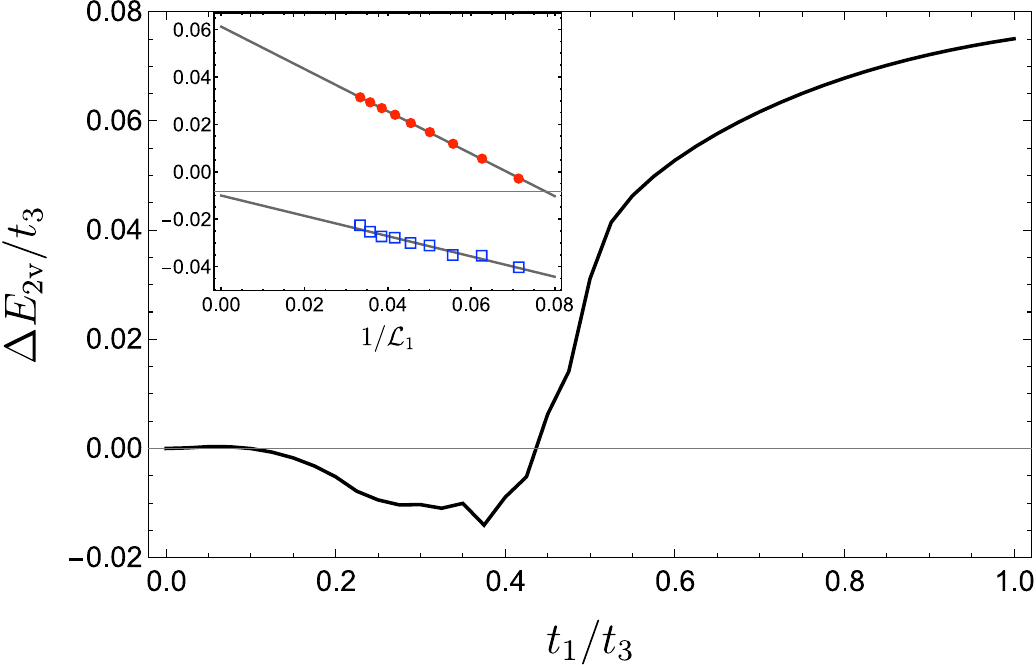}
    \caption{Energy  of a pair of adjacent vortices  in the effective tight-binding model of Eq. (\ref{tbwithuij}). The energy is calculated by extrapolating the result for finite-size systems to the thermodynamic limit, as shown in the inset for $t_1/t_3=0.35$ (blue squares) and $t_1/t_3=0.7$ (red circles). }
    \label{fig5}
\end{figure}

We consider a  state with localized visons created  by reversing the sign of a single link variable $u_{ij}$ on a triangle of the kagome lattice. Since this gauge configuration breaks translational invariance, we  calculate the energy of this state numerically by diagonalizing the tight-binding Hamiltonian on a finite lattice with  size $\mc L_1\times \mc L_2$ along the directions of the vectors $\mb a_1$ and $\mb a_2$ with periodic boundary conditions.  This calculation is performed for system sizes up to $\mc L_1=\mc L_2=30$. By subtracting the energy of the vortex-free ground state and extrapolating the result to $\mc L_1\to\infty$, we obtain an estimate for the energy $E_{\rm 2v}$ of the two-vortex   excitation. Since the $\gamma$ perturbation acts on the $\bar \xi^z$ Majorana fermion as well as on $\bar \xi^x$ and $\bar \xi^y$, we multiply the energy calculated from $H_{\rm tb}^{xy}$ by a factor $3/2$ to account for the contribution from all three spin flavors.  The result is shown in Fig. \ref{fig5}. We find that the finite-size effects are stronger for small $t_1/t_3$, but it is clear that the energy $E_{\rm 2v}$ starts off negative  and becomes positive for larger  $t_1/t_3$.  Since the ratio $t_1/t_3$ increases with $\lambda/v$,  see the     effective hopping parameters   in Fig. \ref{fig3}, we conclude that visons become gapped for sufficiently large $\lambda$. This result  confirms that  the stability of the Majorana Fermi surface state against vison excitations requires moving away from the chiral fixed point of decoupled 1D modes(with $\lambda=0$) and  towards a 2D regime with a significant    backscattering amplitude $\lambda\sim v$. We have also considered the case $t_1<0$, which can accessed by reversing the sign of $\lambda$, but found that $E_{\rm 2v}$ is always negative in this case.

Besides creating vison pairs, the $\gamma$ interaction can make the  visons mobile, lowering their energy. At fixed  $\lambda\sim v$, we expect a quantum phase transition out of the Majorana Fermi surface state as we increase $\gamma$ to the point where visons condense. To understand the conditions on the critical  $\gamma$, recall that this operator is relevant and increases under the RG  flow in the 1D theory. Assuming that the RG flow is cut off at the energy scale $W\sim v/L$ set by the bandwidth of the 2D network model, we replace the bare coupling constant $\gamma$ by the effective coupling $\gamma^*(L)\sim \gamma\,  L^{1-\Delta}$, where $\Delta =3/8$ is the scaling dimension of the spin-1/2 matrix field. The transition must happen when the  effective vison  bandwidth  generated by $\gamma^*$ approaches the  gap obtained for $\gamma=0$. Thus, we expect the gapless CSL phase to extend over the regime \be
\frac{|\gamma^*(L)|}{L} \lesssim  \frac {v}{L}\Rightarrow \frac{1}{L}\gtrsim \left(\frac{|\gamma|}{v}\right)^{8/5}. 
\ee
For fixed $\gamma\neq0$, the gapless CSL becomes unstable in the limit $L\to\infty$, reflecting the instability of the chiral fixed point for a single junction of infinitely long chains \cite{XavierPRB2022}.  On the other hand, this analysis suggests that, even though we started in the limit of long chains with low-energy excitations described by a CFT,  the gapless CSL phase actually becomes more stable if we push the result towards the   physically relevant regime of short chains, with fewer sites in the unit cell.

Let us also comment on the behavior of the staggered part of the spin correlation in the network model.  In contrast with the result for the chiral fixed point discussed in Sec. \ref{section-3}, the staggered part of the correlation does not vanish   in the generic Majorana Fermi surface state with nonzero   $\lambda$ and   $\gamma$. However, since the staggered magnetization involves the spin-1/2 primary field and creates visons, this correlation must decay exponentially with a length scale set by the vison gap. The same can be said about the correlation for the   dimerization operator, which is represented by the trace of the spin-1/2   field \cite{XavierPRB2022}. By contrast, recall that the uniform part of the spin correlation, which only involves gapless fermion excitations, decays as a power law according to Eq. (\ref{correlation}). 
 
  \subsection{Bulk perturbation: mass term}
Let us now consider the mass term in  Eq. (\ref{massterm}). As discussed in Sec. \ref{Yjunctionsec}, this operator destabilizes the  critical spin-1 chain, driving  the transition between  Haldane and dimerized phases.  The corresponding term   on the network can be written as\be
 H_m=im\sum_{\nu,\mb R} \int_0^{L} ds\, \bar \xi^a_{\nu+1,\mb R}(s)\bar \xi^a_{\nu,\mb R}(-s).
 \ee
Focusing on the contribution from the complex fermion in Eq. (\ref{complex-fermions}), we consider \be
 H^{xy}_m=im\sum_{\nu,\mb R} \int_0^{L} ds\, \bar \psi^\dagger_{\nu+1,\mb R}(s)\bar \psi^{\phantom\dagger}_{\nu,\mb R}(-s)+\text{h.c.}\,.
\ee
Since this operator is quadratic in the Majorana fermions, we can analyze its effect in the 2D regime with $\lambda\sim v$ by taking the projection to the effective three-band model. Given that the modes associated with the $n=0$ bands vary smoothly inside the unit cell, a reasonable approximation for the projection is to replace  $\bar \psi^{\phantom\dagger}_{\nu,\mb R}(s)\mapsto \frac1{\sqrt{2L}}f_{\nu,\mb R}$. As a result, we obtain\be
 H^{xy}_m\mapsto im'\sum_{\nu,\mb R}f^\dagger_{\nu+1,\mb R}f^{\phantom\dagger}_{\nu,\mb R}+\text{h.c.},
\ee
with $m'\approx  m/2$. This operator is similar to $t_1$ in Eq. (\ref{Htb}) in the sense that it couples modes with different $\nu$. However, while the hopping parameters in Eq. (\ref{Htb})  are of order $v/L$, the projection of the mass term is independent of $L$. To justify treating this operator as a small perturbation to the low-energy theory governed by $v$ and $\lambda$, with bandwidth $W\sim v/L$, we must impose $|m|L\ll  \lambda, v$. Thus, the relevance of the mass term at the 1D fixed point translates into the fact that the approximation breaks down for any $m\neq0$ in the limit $L\to\infty$. As we discussed for the spin-1/2 operator, keeping a finite $L$ is important to stabilize the gapless spin liquid phase, even if only in a narrow parameter regime.

 We can now add the projection of the mass term to the effective tight-binding model in Eq. (\ref{Htb}).   Since this operator   acts within the unit cell, the result is equivalent to adding a constant matrix to the effective Bloch Hamiltonian [see Eq. (\ref{Bloch})]:\be
\tilde{\mc H}_{\rm eff}(\mb k)= \mc H_{\rm eff}(\mb k)+m'\left(\begin{array}{ccc}
0& -i &i\\
i &0 &-i\\
-i&i&0\end{array}\right).
\ee
Diagonalizing the new Hamiltonian, we find that the threefold degeneracy at $\mb k=0$ is lifted because the Dirac cone formed by the lower and upper bands is gapped out  for $m'\neq 0$.   However, the Fermi surface of the middle band persists along the lines $\mb k\cdot \hat{\mb e}_\nu=0$, as can be promptly  verified by checking that the determinant of $\tilde{\mc H}_{\rm eff}(\mb k)$ still vanishes along these lines for $m'\neq 0$.  Remarkably, the 2D Fermi surface of the spin liquid phase remains stable against the mass term in the limit $|m|L\ll \lambda, v$.  Note that gapping out the Dirac cone at the $\Gamma$ point does not modify the leading behavior of low-energy properties of the gapless spin liquid, which are governed by the Majorana Fermi surface.

 \section{Magnetic field response \label{section 5}}

In this section, we analyze  the effects of an external   magnetic field on the Majorana Fermi surface state.  The Zeeman term for a magnetic field applied along the $z$ spin direction is
\begin{align}
    H_{\text{Z}}=-BS^z_{\rm total}=-B\sum_{j,\alpha,\mb R}S^{z}_{j,\alpha,\mb R}. \label{Hzeeman}
\end{align}
The total magnetization $S^z_{\rm total}$ can be written in terms of the integral of the chiral currents in Eq. (\ref{spinop}). Using the representation in Eq. (\ref{current-majorana}), we obtain\be
    H_{\text{Z}}= iB\sum_{\mb{R}}\sum_{\nu}\int_{-L}^{L}ds\,\bar\xi^{x}_{\nu,\mb{R}}(s)\bar\xi^{y}_{\nu,\mb{R}}(s).
\ee 
Since the magnetic field  only couples to   $\bar \xi^x$ and $\bar \xi^y$, the dispersion relation of the   Majorana fermion $\bar \xi^z$ remains unchanged. In terms of the complex fermion  in Eq. (\ref{complex-fermions}), the Hamiltonian reads  (up to a constant)
\begin{align}
     H_{\text{Z}}= B\sum_{\mb{R}}\sum_{\nu}\int_{-L}^{L}ds\, \bar \psi^{\dagger}_{\nu,\mb{R}}(s)\bar\psi_{\nu,\mb{R}}(s).
\end{align}
Thus, the Zeeman term is equivalent to a chemical potential for the complex fermion. The analogous term in the tight-binding model of Eq. (\ref{Htb}) is \be
H_{\text{Z}}=B\sum_{\nu,\mb{R}} f^{\dagger}_{\nu,\mb{R}}f^{\phantom\dagger}_{\nu,\mb{R}} .
\ee

\begin{figure}[t]
    \centering
    \includegraphics[width=0.95\columnwidth]{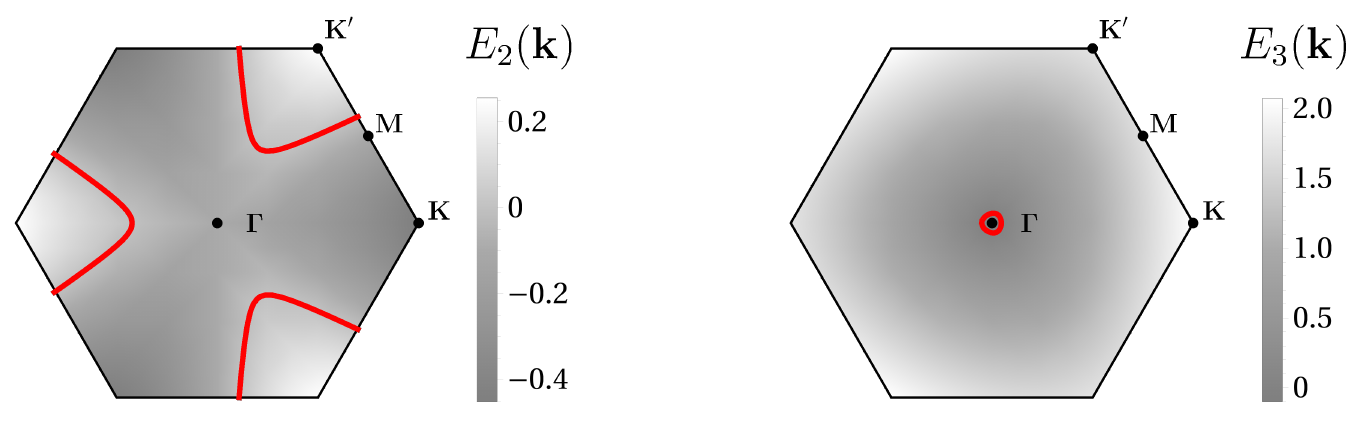}
    \caption{Energy contour plots of the bands $E_{2}(\mb{k})$ (left) and $E_3(\mb k)$ (right)  for  $\lambda/v=0.8$ and $B=-0.1v/L$.   The red lines correspond to the Fermi surface.}
    \label{fig6}
\end{figure}

Since the Zeeman term only shifts the Fermi level for the complex fermion, the spectrum remains gapless for $B\neq0$. However,  the   symmetry $E_r(-\mb k)=-E_{4-r}(\mb k)$ is broken and   the Fermi surface changes significantly; see Fig. \ref{fig6}. The Fermi lines of the middle band $E_{2}(\mb k)$ become curves and no longer cross at the $\Gamma$ point. In addition,  for $B<0$ the upper band $E_3(\mb k)$ crosses the Fermi level and contributes to the Fermi surface with a small pocket around the $\Gamma$ point. Due to the nonzero curvature of the Fermi surface, in the presence of the Zeeman field the equal-time spin correlation for the $S^z$ component decays  as $1/r^3$ \cite{Motrunich2007}. In addition, the magnetic field affects the low-energy single-particle density of states (DOS), $\rho(E)=\frac1{\mc N}\sum_{\mb{k},r}\delta(E_r(\mb k) -E)$. At zero field, we have $\rho(-E)=\rho(E)$, and the DOS has a peak at $E=0$. Since the Zeeman term shifts the Fermi level to $E_F=-B$, the  low-energy DOS  decreases when we turn on the magnetic field. This results, in particular,   in a suppression of  the specific heat $c_V(T)\propto \rho(E_F)T$.

\section{Conclusions \label{conclusions}}

We presented  an analytic approach to study  a 2D gapless phase in  a network built out of    junctions of  spin-1 chains. Gapless phases are hard to explore beyond the approximations of  parton mean-field theory  or even within  coupled-wire constructions  that  assume strong-coupling fixed points at low energies.  By imposing chiral boundary conditions with staggered chirality on the network, we showed that our effective model gives rise to  a phase that shares several low-energy properties with   gapless CSLs found in mean-field approaches on the kagome lattice \cite{ BauerPRB2019, OlivieroSciPost2022}. For instance, this gapless CSL is characterized by a power-law-decaying   spin correlations and a low-energy density of states dominated  by a Fermi surface of spin-1 Majorana fermion excitations.

The main advantage of this approach is   that    fractionalization arises naturally within the effective field theory description of the spin chains. The challenge is to verify that the resulting 2D phase remains stable against perturbations that are formally relevant at the chiral fixed point of the 1D theory. As a key ingredient, the marginal operator associated with backscattering of Majorana fermions at the junctions provides a way to   tune the excitation spectrum along a line of fixed points. Moving along this line to reach  the 2D regime, we were able to associate the spin-1/2 excitations   with gapped visons and to analyze the conditions for  stabilizing the  gapless CSL phase.

Gapless quantum spin liquid states have been proposed for spin-1 systems with bilinear and biquadratic interactions on the triangular lattice, mainly motivated by the material Ba$_3$NiSb$_2$O$_9$  \cite{Bieri2012,Xu2012,Fak2017}. Extrapolating our results to the limit of short chains, we expect that the gapless CSL identified here should be found in spin-1 models on the kagome and star lattices with three-spin interactions. This model could be studied using the same numerical methods that have been applied to the spin-1/2 case \cite{BauerarXiv2013,BauerPRB2019,OlivieroSciPost2022}.  For instance, the effective three-band  model derived here can be used to generate a variational state in a parton representation with spin-1 fermions \cite{LiuPRB2018}. Using variational Monte Carlo, one can compute the energy of this state and compare it against other competing  phases.

Among the directions to be explored in future work, it would be interesting to numerically map out the boundary phase diagram of the junction of spin-1 chains, as done for spin-1/2 chains \cite{BuccheriPRB2018,BuccheriNPB2019}. An accurate quantitative estimate of the location of the chiral fixed point and the critical line defined by the marginal operator would provide guidance for the parameter regime where the CSL phase can be found.   Moreover, a  natural question is whether there are other SU$(N)_{k}$ WZNW models that allow the construction of 2D gapless phases. A lesson from this work is that a  good starting point is to search for  marginal boundary operators in the operator content of the CFT.  Finally, the generalization  to other tricoordinated networks   and   higher spatial dimensions can lead to even more exotic states.

\section{Acknowledgments}
The authors are grateful to Vanuildo  de Carvalho and   Hernan  Xavier for discussions. This work was  supported by a grant from the Simons Foundation (1023171, W.B.F., R.G.P.) and by the Brazilian funding agency  CNPq (F.G.O., R.G.P.). W.B.F. acknowledges FUNPEC under grant 182022/1707. Research at IIP-UFRN is supported by Brazilian ministries MEC and MCTI.

\bibliography{references}
\end{document}